\documentclass{sig-alternate}
\usepackage{mathptmx} 
\usepackage{fancyhdr}
\usepackage[normalem]{ulem}
\usepackage[hyphens]{url}
\usepackage[sort,nocompress]{cite}
\usepackage[final]{microtype}
\usepackage[keeplastbox]{flushend}
\usepackage{amsmath}
\usepackage{amssymb}
\usepackage{amsfonts}
\usepackage{graphicx}
\usepackage{textcomp}
\usepackage{xcolor}
\usepackage{lipsum}
\usepackage{enumitem}
\usepackage{caption}
\usepackage{subcaption}
\usepackage{booktabs}
\usepackage{listings}
\usepackage{float}
\usepackage{makecell}
\usepackage{multirow}
\usepackage{soul}
\usepackage{authblk}
\usepackage[ruled,linesnumbered]{algorithm2e}

\usepackage[bookmarks=true,breaklinks=true,letterpaper=true,colorlinks,linkcolor=black,citecolor=blue,urlcolor=black]{hyperref}

\newcommand{\ir}{IR}
\newcommand{\PROJ}{ZIPPER}
\newcommand{\proj}{\textsc{Zipper}}
\newcommand{\func}{SDE function}
\newcommand{\tsfunc}{sFunction}
\newcommand{\tefunc}{eFunction}
\newcommand{\pdfunc}{dFunction}

\newcommand{\tdys}{classic}

\newcommand{\ttt}[1]{{\texttt{#1}}}
\newcommand{\bench}[1]{{\texttt{#1}}}

\newcommand{\TODO}[1]{{\color{red} {\bf #1}}}

\newcommand{\Sec}[1]{Section~\ref{#1}}
\newcommand{\Tbl}[1]{Table~\ref{#1}}

\newcommand{\Fig}[1]{Figure~\ref{#1}}

\renewcommand{\paragraph}[1]{\vspace*{.1cm}\noindent\textbf{#1}\hspace*{.05cm}}
\newcommand{\PBox}[1]{\vspace*{.05cm}\noindent\fbox{\parbox{\columnwidth}{\vspace*{.05cm}{#1}}}\vspace*{.05cm}}
\soulregister\cite7 
\soulregister\ref7
\soulregister\paragraph7
\soulregister\TODO7
\soulregister\Fig7

\graphicspath{ {./fig/} }

\fancypagestyle{firstpage}{
  \fancyhf{}
  
  \fancyhead[C]{\vspace{15pt}\normalsize{}} 
  \fancyfoot[C]{\thepage}
}

\title{\PROJ{}: Exploiting Tile- and Operator-level Parallelism for General and Scalable Graph Neural Network Acceleration\vspace{-40pt}}

\author{\large\textbf{
Zhihui~Zhang$^1$,~
Jingwen~Leng$^1$,~
Shuwen~Lu$^1$,~
Youshan~Miao$^2$,
}\\ \vspace{-6pt}\textbf{
Yijia~Diao$^1$,~
Minyi~Guo$^1$,~
Chao~Li$^1$,~
Yuhao~Zhu$^3$
}\\\vspace{6pt}
$^{1}$Shanghai Jiao Tong University, $^{2}$Microsoft Research, 
$^{3}$University of Rochester}

\begin{document}
\maketitle
\thispagestyle{firstpage}
\pagestyle{plain}

\begin{abstract}
Graph neural networks (GNNs) start to gain momentum after showing significant performance improvement in a variety of domains including molecular science, recommendation, and transportation.  
Turning such performance improvement of GNNs into practical applications relies on effective and efficient execution, especially for inference. 
However, neither CPU nor GPU can meet these needs if considering both performance and energy efficiency.
That's because accelerating GNNs is challenging due to their excessive memory usage and arbitrary interleaving of diverse operations.
Besides, the semantics gap between the high-level GNN programming model and efficient hardware makes it difficult in accelerating general-domain GNNs.

To address the challenge, we propose \proj{}, an efficient yet general acceleration system for GNNs.
The keys to \proj{} include a graph-native intermediate representation (IR) and the associated compiler.
By capturing GNN primitive operations and representing with GNN IR, \proj{} is able to fit GNN semantics into hardware structure for efficient execution.
The IR also enables GNN-specific optimizations including sparse graph tiling and redundant operation elimination.
We further present an hardware architecture design consisting of dedicated blocks for different primitive operations, along with a run-time scheduler to map a IR program to the hardware blocks.
Our evaluation shows that \proj{} achieves $93.6\times$ speedup and $147\times$ energy reduction over Intel Xeon CPU, and $1.56\times$ speedup and $4.85\times$ energy reduction over NVIDIA V100 GPU on averages.

\end{abstract}

\section{Introduction}\label{sec:intro}

Graph neural networks (GNN) start to gain momentum since researchers involve \textit{graphs} into DNN tasks.
By leveraging the end-to-end and hierarchical learning capability of deep learning, as well as the rich structural information of graphs, 
GNNs achieve better performance in a variety of domains including molecular science~\cite{Molecule}, recommendation~\cite{PinSage,GraphRec}, and transportation~\cite{transportation, Res-RGNN}.  

To better unleash the power of GNNs via efficient inference execution,
we first perform a thorough analysis of the general GNN design space.
We find that GNNs consist of diverse primitive operations, including regular, compute-intensive general matrix multiplication operations from DNNs and irregular, memory-intensive graph operations, such as gather and scatter, from traditional graph processing. 
As such, systems optimize exclusively for DNNs or traditional graph processing are sub-optimal for GNNs. 
Meanwhile, the primitive operations can be freely interleaved in general GNN models. As a result, prior GNN accelerators that focus on a particular kind of GNN~\cite{HyGCN,EnGN,GReTA,GNNerator,ReGraphX} are not generally applicable

Then we build \proj{}, an efficient yet general acceleration system for GNNs.
The fundamental challenge in accelerating the general domain of GNN is the semantics gap between the high-level GNN programming model and efficient hardware. 
Today's GNN programming model~\cite{DGL} is designed to define operations on an input graph as a whole by representing all vertices and edges as tensors. 
We refer to it as \emph{classic GNN programming model}.
Such a programming model makes GNNs similar to conventional CNNs and thus friendly to algorithm designers. But it hides the performance-critical graph structures as well as the vertex- and edge-level operations from flexible execution, losing the opportunity of improving system efficiency.

To bridge the semantics gap, we propose a GNN-aware intermediate representation (IR) and the associated compiler, which together automatically extract the graph-specific semantics (e.g., vertex and edge computational graphs) from \tdys{} GNN programming model. 
The compiler takes a GNN model described in \tdys{} popular GNN frameworks (e.g., DGL~\cite{DGL}) and generates an IR program that will later be mapped to the hardware.
The IR captures primitive operations from GNNs into an IR program, 
with its semantics fitting into the hardware structure for efficient execution.
More importantly, the informative IR program enables our compiler to perform GNN-specific optimizations such as sparse graph tiling and redundant operation elimination.

Coupled with the compiler, we propose a GNN accelerator architecture to execute GNN IR programs.
The accelerator hardware consists of dedicated execution blocks for different primitive operations, along with a run-time scheduler to map an IR program to the hardware blocks.
For better performance, the scheduler effectively exploits GNN-specific parallelisms while respecting the dependencies enforced in an IR program. 
For instance, the scheduler overlaps the execution of tiles (subgraphs) which exercise different hardware resources/blocks to improve hardware utilization.

In evaluation, we compare \proj{} with
Intel Xeon E5-2630 v4 CPU and NVIDIA V100 general-purpose GPU. 
The experiments results show that \proj{} achieves $93.6\times$ speedup and $147\times$ energy reduction over the CPU on average. 
Compared to the GPU, \proj{} achieves $1.56\times$ speedup and $4.85\times$ energy reductions. 

We summarize the main contributions below:
\begin{itemize}[leftmargin=10pt]
    \item We perform a thorough characterization on general GNN models today. Our characterizations show that GNN workloads have a mixed set of compute-intensive and memory-intensive operators that do not simultaneously exist in either traditional graph analytics or DNNs.
    \vspace*{-0.2cm}
    \item We present a GNN IR. It captures primitive operations in GNNs, which is friendly to hardware semantics. The associated compiler automatically converts a GNN model into an IR program while applying GNN-specific optimizations such as tiling and redundant computation elimination.
    \vspace*{-0.2cm}
	\item We propose an efficient and flexible GNN accelerator architecture. The architecture exploits the parallelisms unique to GNN to maximize hardware utilization, thereby improving execution efficiency. The architecture is also applicable to a broad domain of GNNs.
	\vspace*{-0.2cm}
	\item We evaluate \proj{} with detailed experiments and demonstrate average $93.6\times$ and $1.56\times$ speedup with $147\times$ and $4.85\times$ energy reduction over CPU and GPU, respectively.
\end{itemize}

\section{Background}
\label{sec:background}

GNN models form a large design space, and prior work~\cite{HyGCN, AWB-GCN} usually focus on only specific GNN models (e.g., GCN (or graph convolutional network)~\cite{GCN}) and, thus, lack general applicability. Since our work targets generic GNN models, this section describes the general design space of GNN models, emphasizing the common computation primitives.

GNNs extend traditional graph processing with the end-to-end learning capability of deep learning, which has led to better accuracies than the prior hand-crafted or intuition-based methods (e.g., DeepWalk~\cite{DeepWalk} and node2vec~\cite{node2vec}) in a wide variety of domains including molecular science~\cite{Molecule}, recommendation~\cite{PinSage,GraphRec}, and transportation~\cite{transportation, Res-RGNN}. 

Similar to conventional DNNs, GNNs are composed of layers, where a layer $l$ takes as input the vertex and edge embedding matrix along with the graph structure in the form of the adjacency matrix and outputs the new embedding matrix for the layer $l+1$. 
Different from DNNs which may consist of a large number of layers, GNN models only have a few (usually less than five) GNN layers.

\paragraph{Capturing GNN Design Space.}
Owing to the end-to-end learning capability of GNNs, algorithm researchers have explored a wide variety of different GNN models, leading to an enormous model space for GNNs~\cite{GNN:design-space}.

While the GNN design space is vast, computations in GNN can be abstracted as two main kinds of operations~\cite{DGL}: 1) graph operations (\ttt{GOP}) and 2) neural-network operations.
The latter could be further classified into either general matrix multiplication (\ttt{GEMM}) or element-wise (\ttt{ELW}) operations.
These three operations (\ttt{GOP}, \ttt{GEMM}, \ttt{ELW}) cover all forms of computation in GNNs. 
This abstraction is described in a widely-used GNN library \bench{DGL}~\cite{DGL}, and also shared by other libraries such as PyG~\cite{PyG} and NeuGraph~\cite{NeuGraph}.

We now briefly describe these three operations and show how they are used with concrete GNN models.

\paragraph{Graph Operations.}
There are two graph operations in GNNs: \texttt{scatter} and \texttt{gather}. Such two operations can be considered as \emph{vectorized} graph propagation operations in traditional graph processing (c.f., GAS~\cite{GAS} for graph processing).
\begin{itemize}[leftmargin=10pt]
    \vspace*{-0.1cm}
    \item The \texttt{scatter} operation distributes the embedding of each vertex to its outgoing (or incoming) edges.
    \vspace*{-0.2cm}
    \item The \texttt{gather} operation collects and \textit{reduces} the embeddings of all the incoming (or outgoing) edges of each vertex to a fixed-length vertex embedding. The specific \texttt{reduction} function is user-defined, and 
    aligns vertices' embedding for subsequent operations despite different edges.
    \vspace*{-0.2cm}
\end{itemize}

\paragraph{General Matrix Multiplication Operations.}
GNNs use neural network (NN) operations such as MLP\cite{minsky2017perceptrons} and RNN\cite{elman1990finding} to transform vertex and edge embeddings.
These NN-based operations are important to GNNs as they enable the learning ability on the graph.
Since there is no dependency among vertices and edges, these NN operations can be performed in parallel, which is essentially general matrix multiplication.

\paragraph{Element-Wise Operations.}
GNNs also use element-wise (\ttt{ELW}) operations to transform vertex and edge embeddings.
Common \ttt{ELW}s include \texttt{add}, \texttt{exp}, and \texttt{RELU}.
\ttt{ELW} operations on different vertices and edges can also be executed in parallel.
While an ELW operation is less compute-intensive than a \ttt{GEMM}, GNNs can spend a significant portion of time on ELWs owing to their quantity as we show later. 

\begin{figure}[t]
    \centering
    \includegraphics*[trim=1cm 1cm 1cm 0.5cm,width=0.95\linewidth]{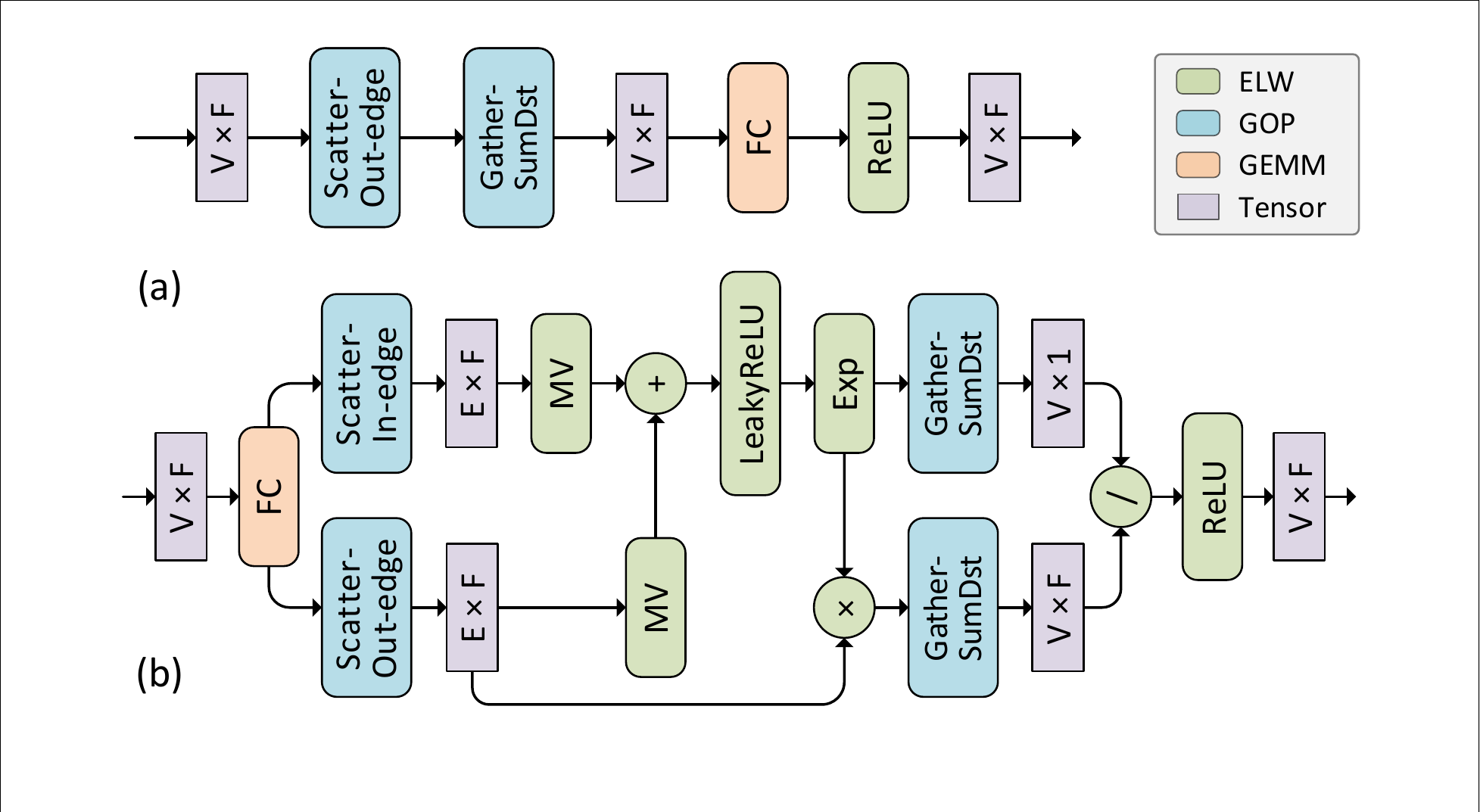}\\
    \vspace{-4pt}
    \caption{One single layer of a)~GCN~\cite{GCN} and b) GAT~\cite{GAT}.}
    \label{fig:gnn-example}
    \vspace{-8pt}
\end{figure}

\paragraph{Examples.}
We take a layer from two popular GNN models, GCN~\cite{GCN} and GAT~\cite{GAT}, respectively, as examples to illustrate how we can express GNNs with such primitive operations.
\Fig{fig:gnn-example} shows each layer's computation graph implemented in the DGL~\cite{DGL} library, where $F$, $V$, and $E$ represent the embedding size, vertex number, and edge number of the input graph, respectively.
 We annotate the three primitive operations in the figure.
The GCN model in \Fig{fig:gnn-example}a show is relatively simple.
In contrast, the GAT model in \Fig{fig:gnn-example}b shows much more complex computation patterns.
The mixed operation types also show the complexity and operation diversity of GNNs.

\section{Motivation}\label{sec:motivation}

A GNN model by nature is a flexible combination of DNNs and graph processing. 
Diverse characteristics of DNNs and graph processing introduce great expressiveness but also make it challenging for accelerating computation.
In this section, we first study the characteristics of the GNN workloads against the well-studied DNN models and traditional graph processing algorithms on GPU architecture.
We find that the existing solutions suffer from inefficiency problems due to GNN's diverse characteristics.
We then study the root cause at both architecture and software level,
and propose a 
solution with co-design of software and hardware.

\begin{figure}[t]
    \centering
    \vspace{-1pt}
        \hspace{-6pt}
        \includegraphics*[trim=0.5cm -0.2cm 1cm -0.2cm,width=0.98\linewidth]{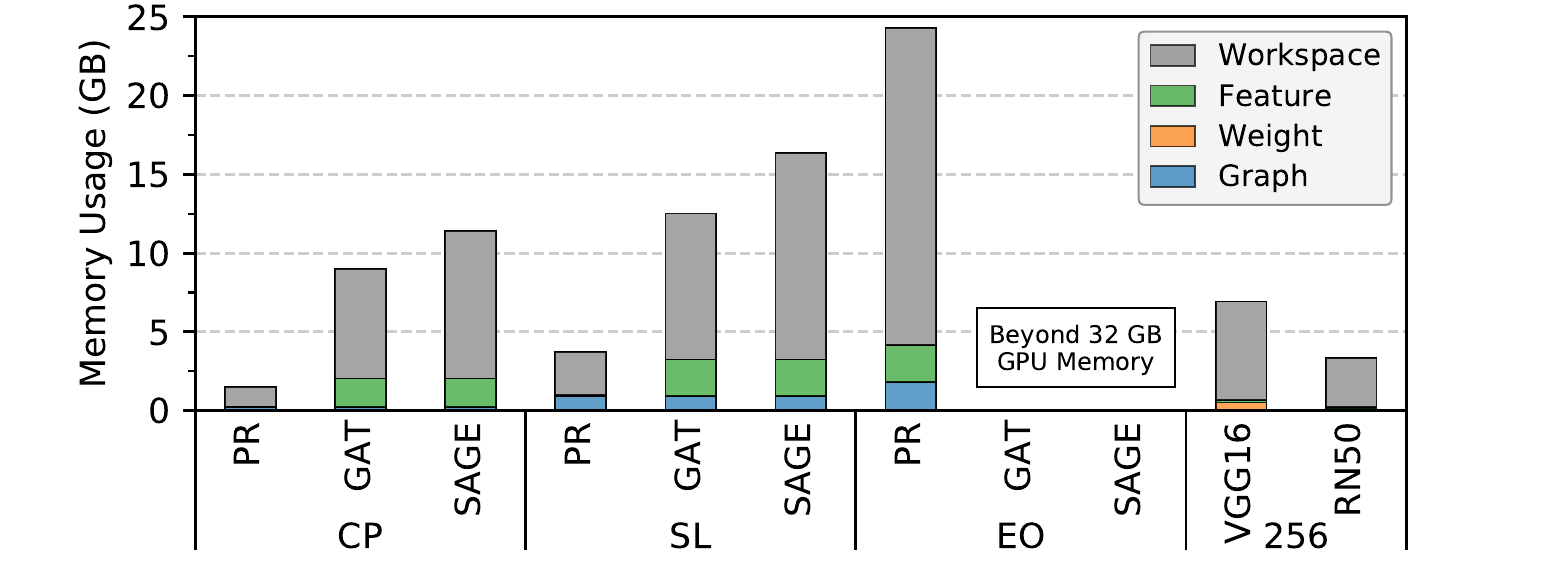}
        \hspace{-6pt}\\
        \vspace{-12pt}
        \caption{The total memory usage. The workspace refers to the intermediate data during the computation. GNNs (\bench{GAT} and \bench{SAGE}) require much larger workspace and thus could not process large graphs (e.g., EO).}
        \label{subfig:total_memory_usage}
    \vspace{-15pt}
\end{figure}

\begin{figure*}[t]
    \centering
    \vspace{-6pt}
    \hspace{-9pt}
    \includegraphics*[trim=0cm 0.3cm 0.2cm 0cm,width=0.25\linewidth]{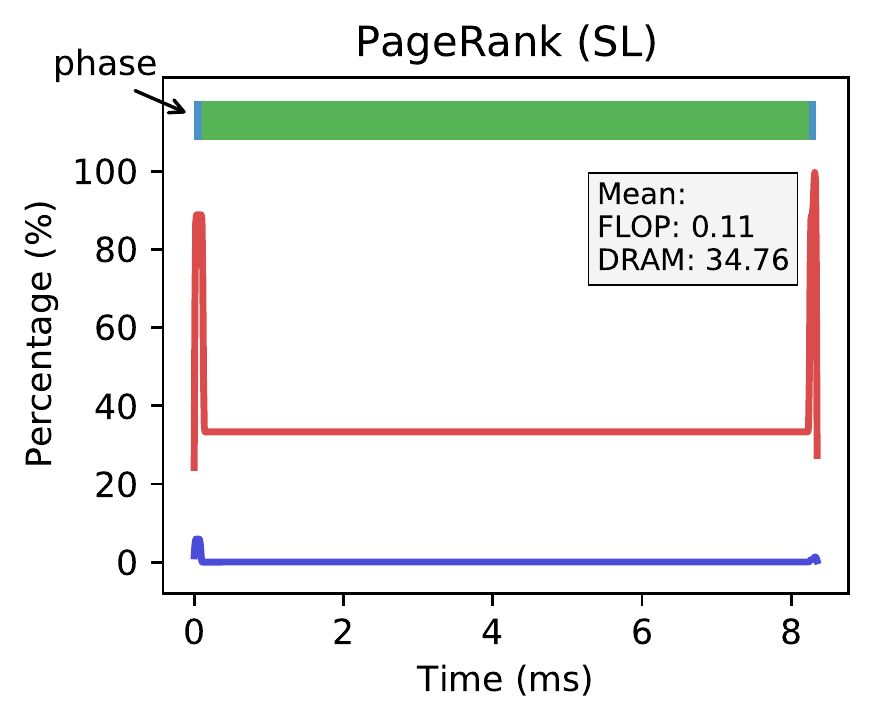}\hspace{-2pt}
    \includegraphics*[trim=0cm 0.3cm 0.2cm 0cm,width=0.25\linewidth]{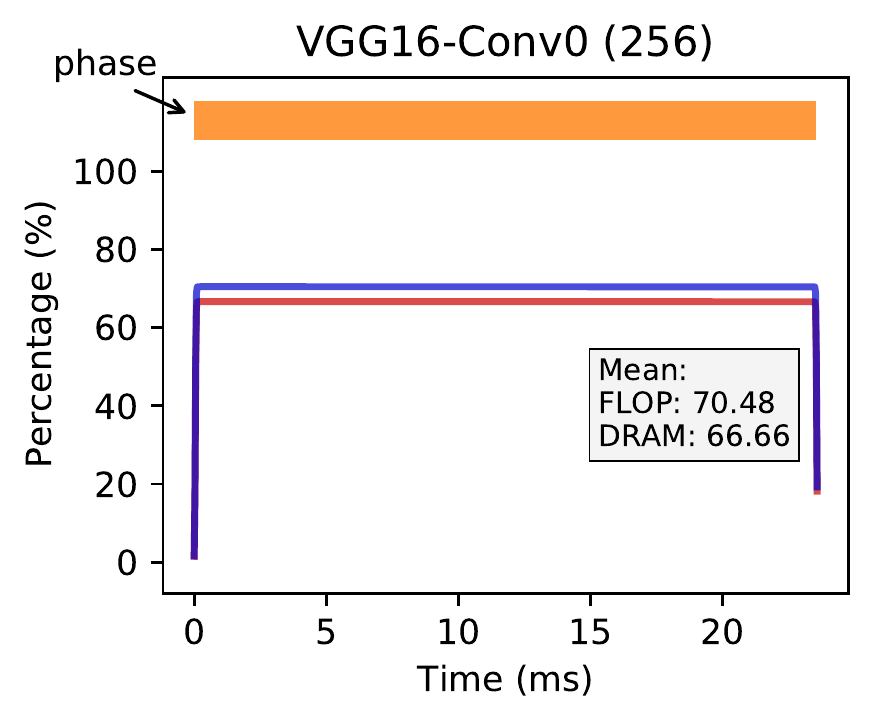}\hspace{-2pt}
    \includegraphics*[trim=0cm 0.3cm 0.2cm 0cm,width=0.25\linewidth]{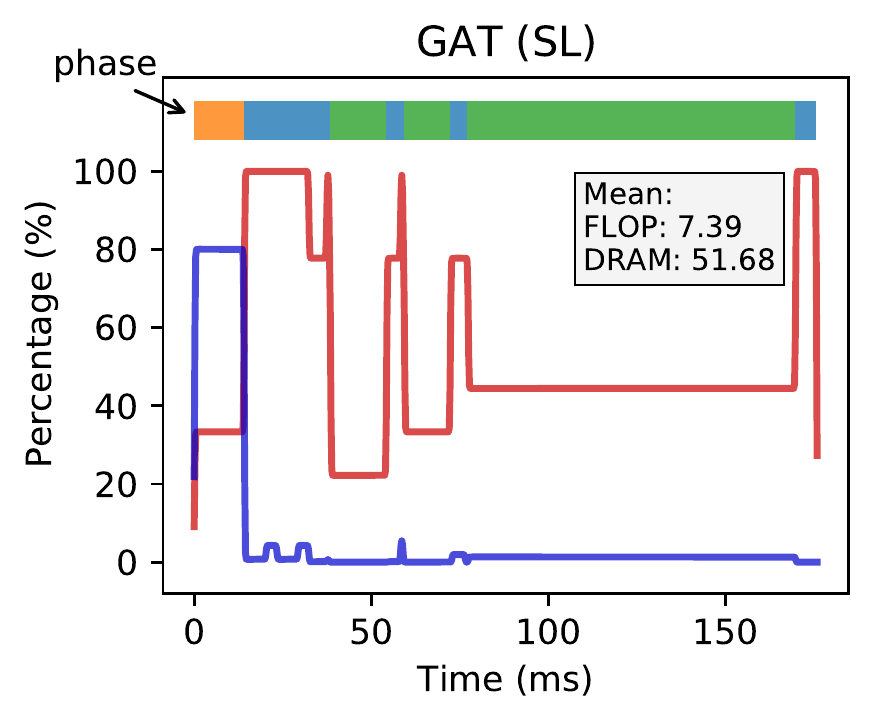}\hspace{-2pt}
    \includegraphics*[trim=0cm 0.3cm 0.2cm 0cm,width=0.25\linewidth]{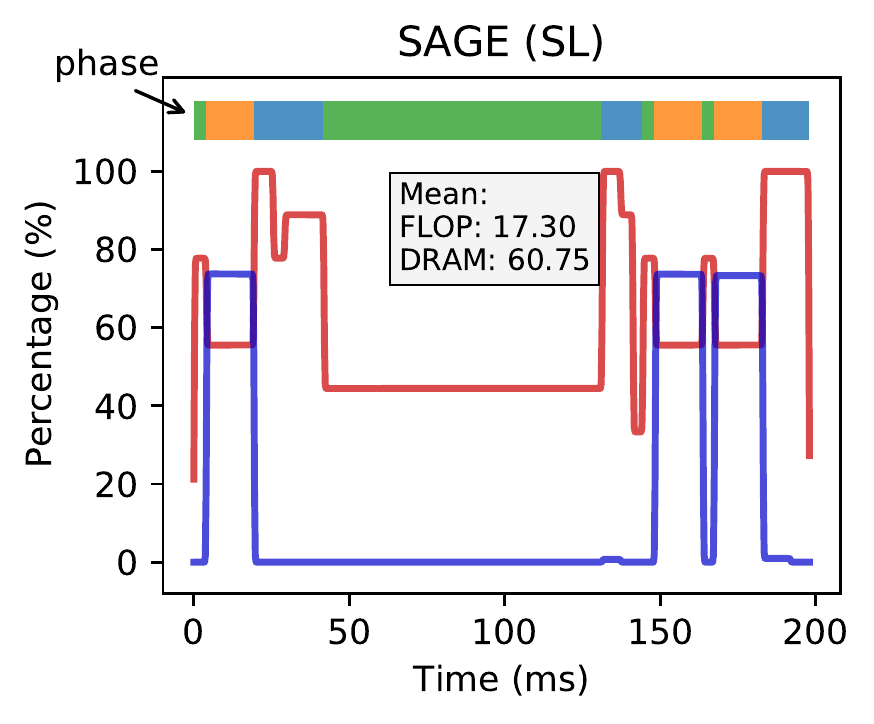}\hspace{-10pt}\\
    \hspace{5pt}\includegraphics*[trim=0.52cm 0.37cm 0.5cm 0.37cm,width=0.98\linewidth]{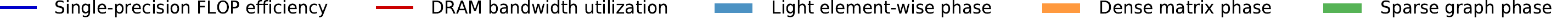}
    \vspace{-2pt}
    \caption{The metric kernel traces of one layer or iteration. We also annotate different phases.}
    \label{subfig:metric_trace}
    \vspace{-8pt}
\end{figure*}

\subsection{GNN Workload Characterization}\label{subsec:characterization}

We compare and contrast GNNs against DNNs as well as traditional graph processing algorithms to understand GNNs' common and unique properties.
Specifically, we start our study from two representative GNN models, \bench{GAT}~\cite{GAT} and \bench{GraphSAGE} (\bench{SAGE})~\cite{GraphSAGE}, two DNN models, \bench{VGG}~\cite{VGG} and \bench{ResNet} (\bench{RN})~\cite{ResNet}, and a traditional graph processing workload \bench{PageRank} (\bench{PR})~\cite{PageRank}. GNNs and traditional graph processing algorithm are evaluated on three datasets (\Tbl{tbl:benchmark:dataset}); DNNs are evaluated on ImageNet~\cite{ImageNet} with batch size of 256.

We identify two opportunities for optimizing GNNs.

\PBox{%
\textbf{Observation 1:} \textit{GNNs exhibit much higher memory usage than DNNs and traditional graph processing. As a result, 
it's hard to efficiently scale GNNs to large graphs.}
}

Specifically, \Fig{subfig:total_memory_usage} compares the memory usage of different algorithms on a NVIDIA V100 GPU with 32~GB memory. On the dataset SL, the GNN \bench{GraphSAGE} uses 16.3~GB of GPU memory, while \bench{PageRank} on the same dataset uses only 3.7~GB and \bench{VGG16} with ImageNet under batch size of 256 uses 6.9~GB. 
We also observe a similar trend on dataset CP.

The excessive memory usage prevents GNN models from processing large graphs. \Fig{subfig:total_memory_usage} shows that when using a large graph EO, which consists of $10.5\times$ more vertices and $1.2\times$ more edges than SL, both GNNs \bench{GAT} and \bench{GraphSAGE} run into the out-of-memory issue.

To understand the excessive memory usage, we further break down the memory usage into four components: the graph data, the weight matrices, the input/output feature embeddings, and the workspace, which refers to the intermediate data between operations. We observe that GNNs use a significant portion of memory for storing the intermediate data. This is fundamental because \tdys{} GNN systems operate on an entire graph within one operation.

\PBox{%
\textbf{Observation 2:} \textit{Due to irregular nature of graph, similar to traditional graph processing, GNNs also show lower hardware utilization than CNNs. However, GNNs mix diverse primitive operations, which requires different types of hardware resources. It provides a unique optimization opportunity by redistributing the hardware resource while overlapping different operations.
}
}%

\paragraph{Primitive Operation Diversity} \Fig{subfig:metric_trace} plots how the single-precision FLOP efficiency and the DRAM bandwidth utilization change over time for the four algorithms -- all on a V100 GPU.
These two metrics capture key computational and memory behaviors of an algorithm. The data is obtained from one iteration of \bench{PageRank} and a layer in the GNN and DNN models. At the top of each figure, we annotate which primitives operation (i.e., \ttt{GEMM}, \ttt{ELW}, and \ttt{GOP} in \Sec{sec:background}) dominates an algorithm at a given moment in time.

GNNs exhibit a much more diverse mix of primitive operations than traditional graph processing and CNNs. GOP dominates the execution of \bench{PageRank}, while \ttt{GEMM} and \ttt{ELW} dominate the execution of \bench{VGG}.
In contrast, all the three primitives operations exist in GNNs; the interleaving of the three operations varies by GNN model.

It is worth noting that prior GNN accelerators~\cite{HyGCN,GraphACT,EnGN,GReTA,AWB-GCN} target only one particular type of GNN called GCN (graph convolutional network), which has a fixed primitive operation interleaving of \texttt{GOP}-\texttt{GEMM}-\texttt{ELW} as shown in \Fig{fig:gnn-example}a. 
A general GNN would be much flexible than GCN in model structure, as \texttt{GAT} shown in \Fig{fig:gnn-example}b, which has a much more complicated primitive operation mixing and interleaving.

\paragraph{Low Hardware Utilization} Because of the primitive operation diversity, GNNs tend to have lower hardware utilization than CNNs. As a result, accelerators built for CNNs are ill-suited for GNNs. We annotate in \Fig{subfig:metric_trace} the average FLOP efficiency and DRAM bandwidth utilization for each algorithm. The FLOP efficiency on both GNNs is at least 35\% lower than that of \bench{VGG}; the DRAM bandwidth utilization of both GNNs is also lower than that of \bench{VGG}.

A close examination of the GNN execution shows the reason. CNNs primarily rely on \ttt{GEMM}, which has high FLOP efficiency and high DRAM bandwidth utilization due to its regular compute and memory access patterns. GNNs mix regular \ttt{GEMM} kernels and irregular \ttt{GOP}s used in traditional graph processing (e.g., \bench{PR}), which has low FLOP efficiency and DRAM bandwidth utilization due to its irregular compute and memory access patterns. As a result, GNNs tend to have lower hardware utilization than CNNs.

\begin{figure}[t]
    \centering
        \includegraphics*[width=1\linewidth]{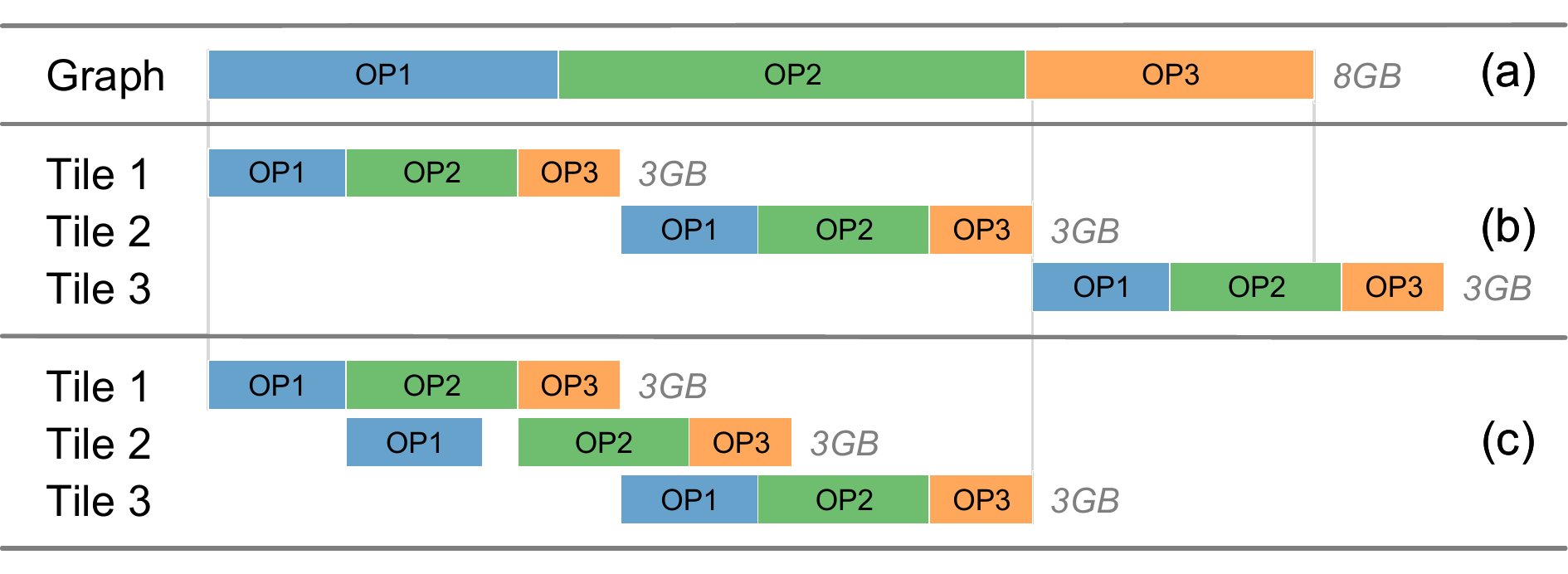}
        \caption{An illustration of the benefits of inter-tile pipelined execution. (a) Non-tiling execution with the peak memory usage of 8~GB; (b) tiling-based execution, where the graph is partitioned into three tiles and the memory usage is 3~GB; (c) inter-tile pipelined execution, where two tiles are pipelined in the operation level with no more than 6~GB memory usage.}
        \label{fig:pipeline}
    \vspace{-4pt}
\end{figure}

\subsection{Inter-tile Pipelining}

Our main idea is to exploit the operation diversity in GNNs 
and pipeline the operations
to reduce memory footprint while improving hardware utilization. \Fig{fig:pipeline} illustrates the idea of a simple GNN with three primitives operations.

\Fig{fig:pipeline}a illustrates how GNNs are executed on today's system, where the three operations are sequentially executed as three serialized stages, each operating on \textit{the entire graph}. The intermediate data between stages encodes information for the entire graph, leading to a large memory footprint. In addition, the serialized execution leads to long execution times with low hardware resource utilization.

A common strategy to reduce the memory footprint is graph tiling~\cite{GridGraph}, which partitions a graph into smaller subgraphs, a.k.a., tiles, and operates on each tile separately. \Fig{fig:pipeline}b illustrates such an idea, where the entire graph is divided into three tiles. The three tiles are processed sequentially, which reduces the memory footprint since at any given moment only a small subgraph is resident in memory.

However, this strategy degrades performance due to the bookkeeping overhead such as setting up and switching tiles. In addition, it does not address the low hardware utilization, as at any given moment only one operation is executed.

We propose to pipeline across tiles, as illustrated in \Fig{fig:pipeline}c, which retains the advantage of low memory footprint while significantly improving the performance. By pipelining across tiles, operations of different tiles are overlapped and executed at the same time. Overlapping operations exercise different resources, improving the overall resource utilization
and therefore leads to better performance.

\subsection{Challenges of Inter-tile Pipelining}\label{subsec:challange}

\begin{figure}[t]
    \centering
        \includegraphics*[trim=0.1cm 0cm 0cm 0cm,width=1\linewidth]{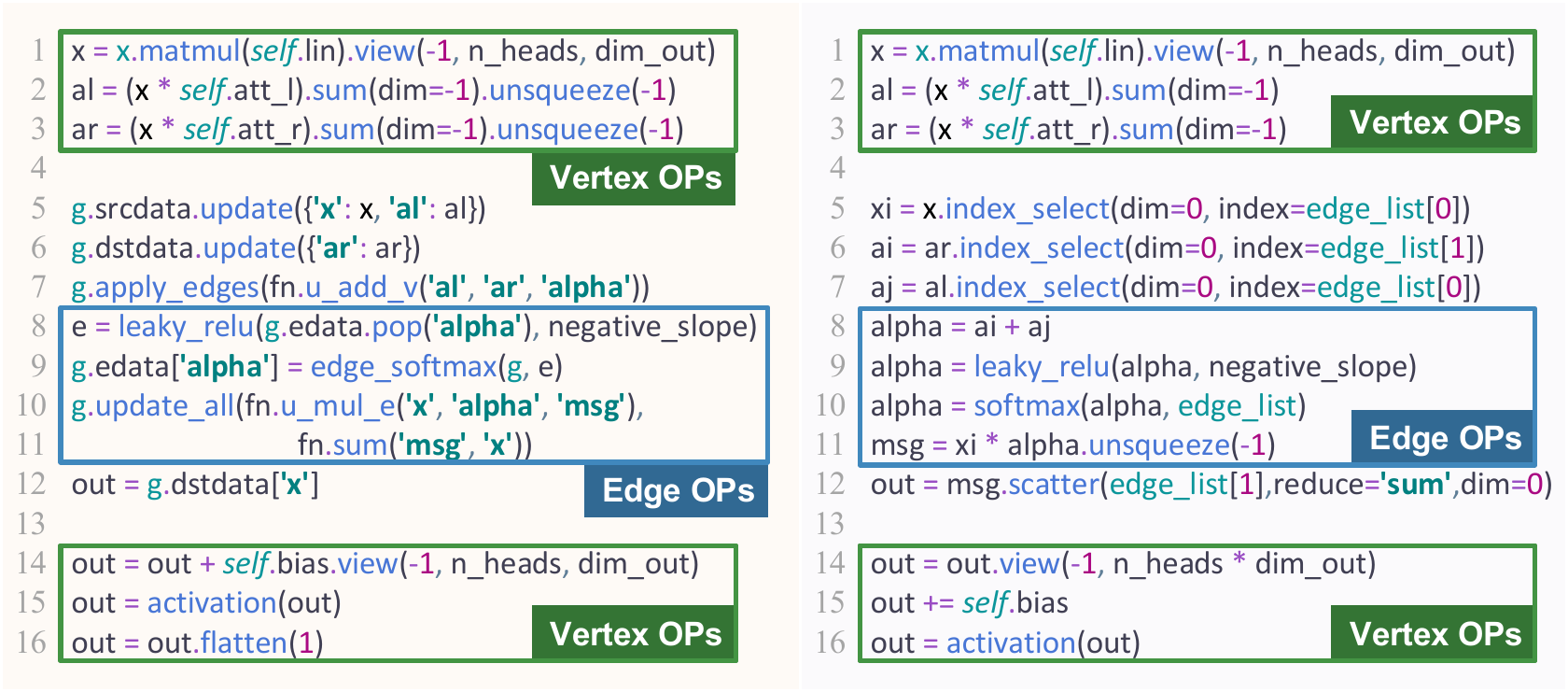}\\
        \caption{The unawareness of graph semantics in current GNN frameworks: DGL~\cite{DGL} (left) and PyG~\cite{PyG} (right).}
        \label{fig:challenge}
    \vspace{-4pt}
\end{figure}

Applying tile-level pipelining to GNNs is challenging for two reasons.
First, there is a semantics gap between the \tdys{} GNN programming model and efficient hardware execution. In particular, tile-level pipelining requires us to identify the operations associated with each tile, including its vertices and edges. But \tdys{} GNN programming model is designed to define operations on a graph as a whole without exposing vertex- and edge-level operations. 
This is accomplished by representing all vertices and edges as tensors. This programming model thus expresses GNN execution as tensor computations, similar to conventional CNNs. 
\Fig{fig:challenge} shows such an example from the GNN library in PyTorch, where the bold boxes show how the vertices and edges are represented as tensors and the GNN execution is represented as tensor computation without exposing graph semantics.

While this programming model makes GNNs similar to conventional CNNs and is thus friendly to GNN algorithm designers, it also hides vertex and edge-level details that are vital for efficient hardware execution.

Second, GNNs adopt a wide variety of different operations. Therefore, one must flexibly schedule a GNN to the hardware. 
For instance, the \texttt{GCN} and the \texttt{GAT} in \Fig{fig:gnn-example} are drastically different. A static, fixed mapping from a GNN to the hardware is likely suboptimal in utilizing the hardware.

\begin{figure*}[t]
    \centering
    \includegraphics*[width=0.99\linewidth]{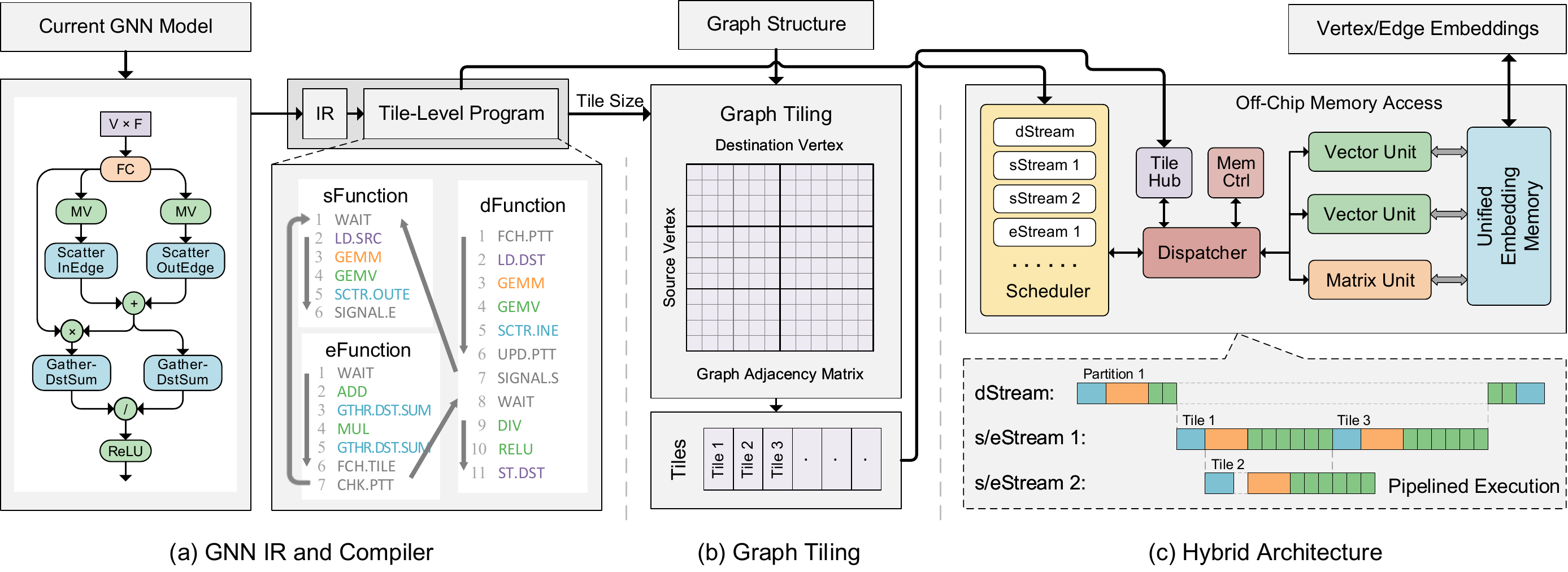}
    \vspace{-4pt}
    \caption{The \proj{} system overview.}
    \label{fig:overview}
    \vspace{-8pt}
\end{figure*}

\section{\PROJ{} Overview}
\label{sec:base}

In this work, we propose a hardware and software co-designed system that exploits the tile- and operator-level parallelism to provide efficient and scalable support for generic GNN acceleration. 
\proj{} overcomes the challenges of inter-tile pipelining in \Sec{subsec:challange} through a combination of 
software GNN compiling and hardware GNN architecture.
\Fig{fig:overview} shows an overview of the \proj{} system.

First, \proj{} proposes a GNN intermediate representation (IR) and the associated compiler, which together automatically extract the graph-specific semantics (e.g., vertex and edge computation graphs) from \tdys{} GNN programming model. 
The IR is closer to the hardware, and enables an efficient hardware accelerator design and scheduling. 

Second, \proj{} proposes an accelerator architecture for executing GNN models represented as IR programs. The key to the hardware is to be flexible enough to accommodate different types and mixes of the primitive operations while being efficient by exploiting GNN-specific parallelisms and locality. The hardware achieves this by employing dedicated blocks for each primitive operation coupled with an efficient run-time scheduler, which maps GNN IR programs to the hardware substrate.

\paragraph{IR and Compiler}
The proposed IR is structured as multiple directed acyclic graphs (DAG) extracted from a GNN model.
Each graph is labeled as a vertex segment or a edge segment, where the DAG nodes are the GNN operations for a single vertex or an edge while the DAG edges are the data of the vertex or edge.
The IRs are meant to be used by the compiler, which compiles the GNN model into a low- and tile-level program consisting of three functions for the source vertices of tiles (\tsfunc{}), the edges of the tiles (\tefunc{}), and the destination vertex of partitions (\pdfunc{}), respectively, under the tiling-based execution semantics to specifies the tile data dimensions as well as the interactions between the vertices and edges.
We refer to the three functions as \func{}s for simplicity.

\paragraph{Hardware Architecture} The hardware consists of two main components: the building blocks to support various primitive operations execution and a scheduler that dispatches GNN tiles to the hardware blocks.

The hardware blocks consist of both generic matrix and vector units to support \texttt{GEMM} and \texttt{ELM} operations, respectively, as well as graph-specific structures that are optimized for \texttt{GOP}. 
The scheduler generates independent work units for source vertices and edges in tiles and destination vertices in partitions, which we call sStreams, eStreams and dStreams.
The streams are then pipelined efficiently across the hardware blocks to achieve high utilization.

\section{GNN Parallelization}
\label{subsec:tiling}

In order to exploit the inter-tile pipelined execution opportunity, we first describe our graph tiling method and a generally applicable strategy to parallelize
GNN models.
 
\subsection{Graph Tiling}

The fundamental building block in our proposed system is graph tiling (also called partition or sharding), which divides an input graph into smaller sub-graphs, a.k.a., tiles.
The benefits of graph tiling are two-fold: easing the pressure of excessive memory footprint of GNN computation and 
exposing tile-level parallelism that \proj{} leverages (\Fig{fig:pipeline}).

\begin{figure}[t]
    \centering
    \includegraphics*[trim=0.1cm 0cm 0.6cm 0cm,width=1\linewidth]{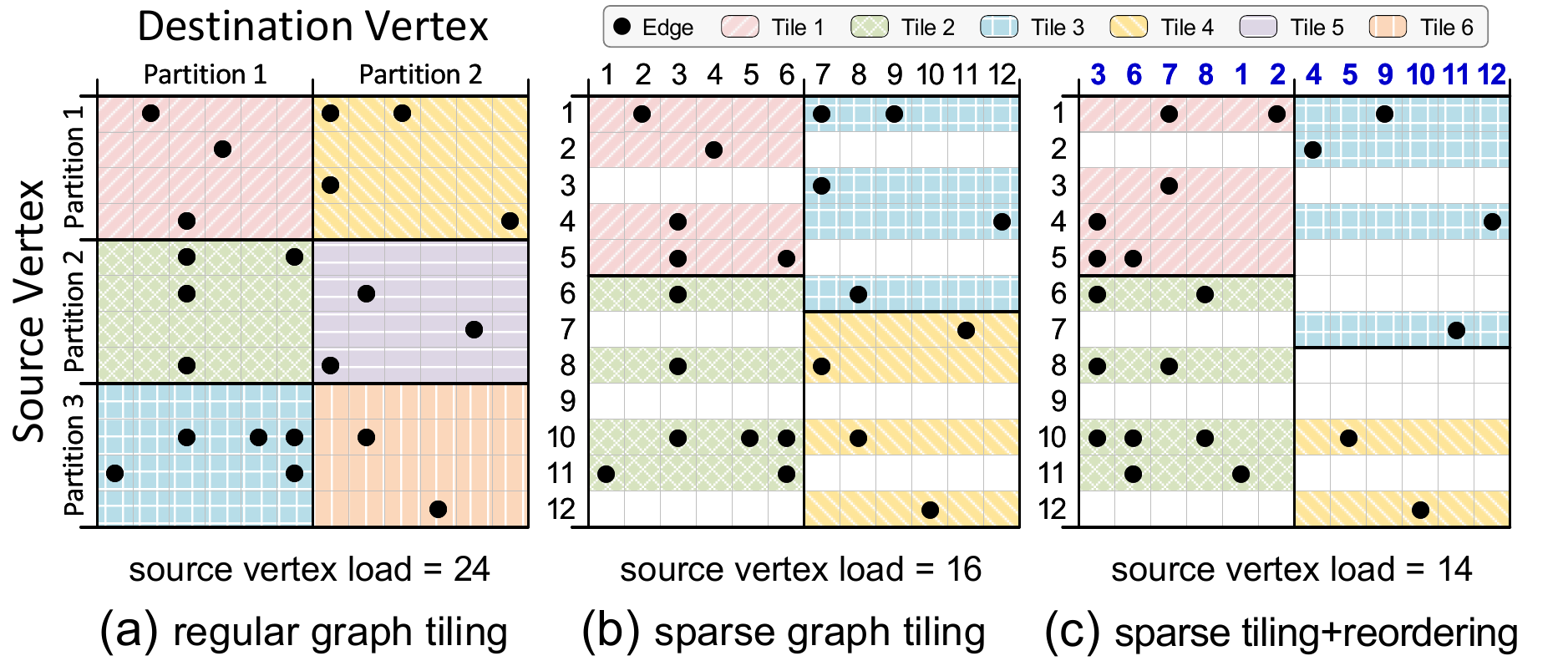}
    \caption{Comparison of different graph tiling methods. The vertices in (c) are arranged in descending order of their in-degrees.  
    The maximum of the source and destination vertices in a tile is 4 and 6 respectively. 
    }
    \label{fig:tilings}
    \vspace{-13pt}
\end{figure}

We adopt a tiling strategy called grid-based or regular tiling~\cite{GridGraph,NeuGraph,GReTA}.
The idea is to divide the graph adjacency matrix into multiple smaller rectangles as tiles.
\Fig{fig:tilings}b illustrates the grid-based tiling example.
Formally, we first split the vertices evenly into several \emph{destination partitions} according to their vertex IDs.
For each destination partition, we further split its vertices into several \emph{source partitions} according to their vertex IDs.
As a result, each tile corresponds to exactly one destination and one source partition and uniquely identifies a set of edges whose source and destination vertices are in the corresponding partitions.

Owing to the sparsity in the graph, the adjacency matrix for the tile that indicates connections between edges and vertices is stored in a sparse format such as edge list (COO) or compressed sparse column (CSC) for saving the storage~\cite{CSC}. 
While the tile metadata such as the edge and vertex numbers are usually stored in a dense array.

\subsection{Multi-streamed Execution}\label{subsec:multistream}

Based on the above grid-based graph tiling, we propose a generally applicable parallel execution mechanism for GNN models which are combinations of \texttt{GEMM}, \texttt{ELM}, and \texttt{GOP}s.

We first need a multi-streamed parallel execution mechanism that supports fine-grained inter-stream synchronization via a signal-wait pair.
The idea is to map the computation of concurrent tiles to different streams and insert proper synchronization instructions to maintain the dependency from the original GNN model.

The key for this mapping is the \texttt{Gather} operation.
This operation is applied to all the tiles under the same partition to reduce the embeddings of all the edges and source vertices for each destination vertex.
As such, all the operations that depend on the result of the \texttt{Gather} operation need to wait for all tiles in the same partition.
In the meantime, the operations that do not depend on the \texttt{Gather} operation can be executed in parallel.
 
Following the above principle, we propose to use multiple streams for the concurrent tile computation and a single stream for the partition computation.
We call the former as \emph{sStreams} and \emph{eStreams}, processing the source vertices and edges in tiles respectively, and the latter as \emph{dStream} for the destination vertices in partitions.
The reason that the number of s/eStreams is greater than the number of dStream is that there can be many tiles for the same partition.

\Fig{fig:overview}c-bottom shows an example of streams.
The dStream first processes the embedding of a partition until it reaches the \bench{signal} and \bench{wait} instructions.
The \texttt{signal} instruction wakes up two sStreams, which starts to load and process a tile under the partition.
The sStream ends up with a \texttt{signal} to wake up an eStreams and continue the tile process.
When the eStream reaches the end, it fetches the metadata of the next tile and checks whether the destination partition is the current one.
If true, the eStream starts a new round of s/eStreams through the \bench{signal} and \bench{wait} instruction; or it resumes the dStream.  
When the dStream continues, it finishes the partition process and fetches the next partition.

\subsection{Tile Parameter Optimization}

Although the tiling-based pipelining reduces the memory footprint of the GNN computation, the large amount of tile vertex data loaded and computed without contributions to the final results because of the graph sparsity.
\Fig{fig:tilings}a illustrates this problem where the source vertex without an edge in a tile causes redundant memory access and computation.
For example, in tile~1 (the top left one), vertex~3 is included in the tile and thus loaded to the on-chip memory.
However, this vertex will be only computed as an edge source and the result will not be propagated to any edges or destination vertices because it does not have any associated edges.
So the on-chip memory load, as well as the associated computation of such vertices, are unnecessary in this tile.

\paragraph{Sparse Tiling.}
We use the sparse graph tiling approach~\cite{NeuGraph} that embraces the graph sparsity to remove the above unnecessary processing. 
As illustrated in \Fig{fig:tilings}b,
only the source vertices that have associated edges in the tile are kept.
Usually, traditional graph processing systems, e.g.~\cite{X-Stream}, prefer regular tiling to sparse tiling. 
It is because their vertex/edge features are scalars. Applying sparse tiling would turning efficient sequential off-chip memory access into much more expensive random access on small-size scalars, leading to lower overall performance even skipping unnecessary vertices.
However, the vertex/edge features in GNN are high-dimensional embeddings, much larger than the scalars and each of them can be translated into multiple off-chip memory transactions,
which can match the bandwidth of sequential access while avoiding unnecessary processing.

\paragraph{Graph Reordering.}
The improvement of the sparse tiling is limited yet because of the random vertex distribution.
As shown in \Fig{fig:tilings}b, we can observe that most source vertices only have a few (mostly one) edges to the destination partitions in a tile, which less than their out-degrees.
As such, there is an opportunity to gather the out-edges of the source vertices into a few tiles to further reduce the redundancy and improve vertex data reuse.

We leverage graph reordering to fulfill such an opportunity, which makes minimal modifications to the existing system design.
There are lines of sophisticated graph reordering in traditional graph processing, but only the lightweight methods are effective considering the limited performance improvement~\cite{Reorder:BalajiL18,Reorder:DBG}.
In contract, applying the reordering to GNN can lead to considerable performance improvement, because the data volume and computation for a vertex is much more than that in traditional graph processing.
We use a heuristic Degree Sorting strategy that reorders the vertices according to their in-degrees as shown in \Fig{fig:tilings}c to illustrate the effectiveness of reordering for GNN.
The vertices with high in-degrees are arranged to the left side so there are more blank rows on the right side for the sparse tiling.
As a result, the total source vertex load is reduced.

\section{GNN IR and Compiling}

{\renewcommand{\arraystretch}{1.2}
\begin{table}[t]
\caption{An example of the Graph-Native GNN IR operations.}
\label{tbl:ir}
\centering\selectfont
\resizebox{\linewidth}{!}{
\begin{tabular}{clll}
\toprule
\multicolumn{1}{c}{\textbf{Operation Type}}
& \multicolumn{1}{c}{\textbf{Examples}}
& \multicolumn{1}{c}{\textbf{Operands}}
\\ \midrule

Computational            
& \begin{tabular}[c]{@{}l@{}}\textbf{GEMM} (\ttt{matmul}, \ttt{conv}), \textbf{ELW} (\\\ttt{add}, \ttt{sub}, \ttt{mul}, \ttt{div}, \ttt{mv}, \ttt{relu}, \ttt{exp})\end{tabular}   
& \begin{tabular}[c]{@{}l@{}}a single edge / vertex\end{tabular} 
\\ \midrule

\begin{tabular}[c]{@{}c@{}} Communicational\\(\textbf{GOP}) \end{tabular}       
& \begin{tabular}[c]{@{}l@{}}\textbf{Scatter} (\ttt{sendOutEdge-recvSrc},\\ \ttt{sendInEdge-recvDst}), \textbf{Gather} (\\\ttt{sendDstSum-recvInEdge})\end{tabular}   
& \begin{tabular}[c]{@{}l@{}}an communication index\\ and an edge / vertex\end{tabular}

\\ \midrule

\begin{tabular}[c]{@{}c@{}} Entry \& Exit \end{tabular}       
& \begin{tabular}[c]{@{}l@{}}\textbf{Indicator} (\ttt{input}, \ttt{output})\end{tabular}   
& \begin{tabular}[c]{@{}l@{}}a string \textit{edge} / \textit{vertex}\end{tabular}

\\ \bottomrule
\end{tabular}}
\end{table}}

\begin{figure*}[t]
    \centering
    \vspace{-2pt}
    \includegraphics*[width=0.99\linewidth]{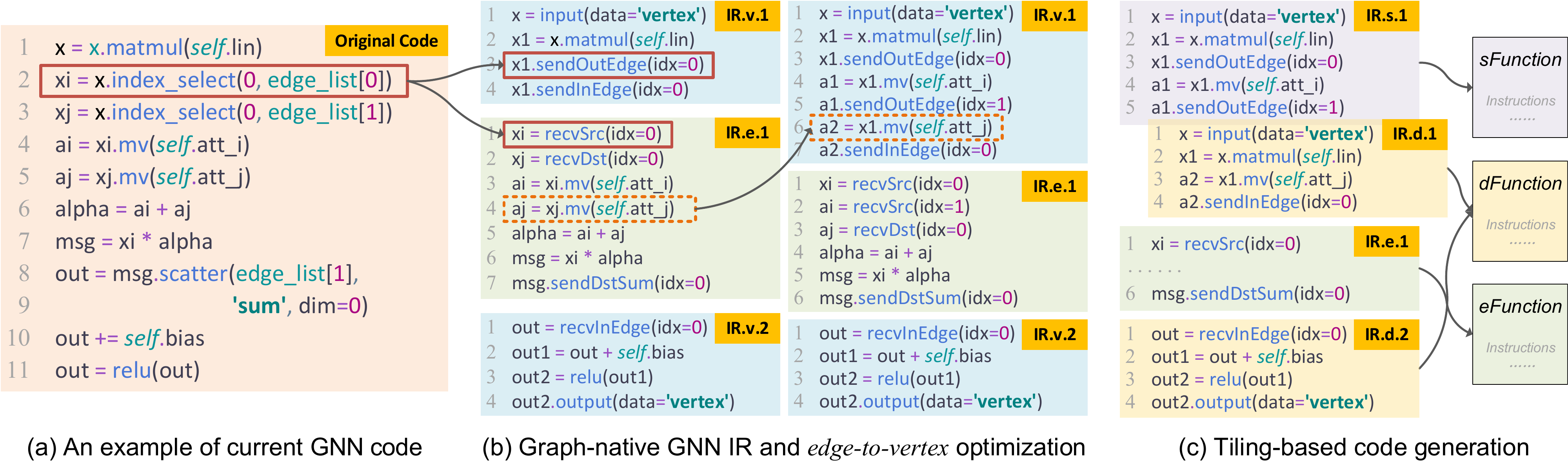}\\
    \vspace{-1pt}
    \caption{Compiling process from a high-level GNN programming model (PyG~\cite{PyG}) to the functions for tiling-based execution.}
    \label{fig:compiling}
    \vspace{-6pt}
\end{figure*}

\proj{} automatically compiles a high-level GNN model expressed in 
\tdys{}
GNN programming framework (e.g., DGL~\cite{DGL}) to the tile-level program that is scheduled and executed as streams on our hardware substrate.
As we have explained earlier, the \tdys{} GNN programming model is centered around tensor computation and does not expose edge and vertex level operations. 
To bridge the semantics gap, we propose a graph-semantics-preserving IR that is used by the compiler to generate the \func{}s mentioned in \Sec{subsec:tiling}.
An end-to-end illustration of the compiling process is depicted in \Fig{fig:compiling}, and we first introduce the \ir{} with the corresponding compiling process, and then discuss performance optimizations based on the \ir{}.

\subsection{Graph-Native GNN IR}

We propose Graph-Native GNN IR that contains multiple computational DAG segments.
Each segment is labeled as an edge or vertex segment, and consists of IR operations as nodes that operate the data of a single edge or vertex.
\Fig{fig:compiling}b shows two examples.

\Tbl{tbl:ir} lists a subset of the IR operations in \proj{}, which are designed to represent all three types of primitive operations (\Sec{sec:background}).
In particular, the computational operations correspond to the \texttt{GEMM} and \texttt{ELW} operations in general DNNs, while the communicational operations correspond to the \bench{GOPs} that exchange data between vertex and edge.
All the IR operations target only one item (i.e., single edge or vertex) for graph semantic atomicity.
The IR connects the high-level programming model by using compatible computational graphs and operations in the DNN, while it recovers the graph semantics by operating on a single edge and vertex with decoupled computational graphs.

\paragraph{Compiler.}
We build a compiler to automatically translate the existing GNNs expressed via the high-level programming model to our proposed IR.
The compiler then uses the generated IR to produce the \func{}s, which are eventually mapped to the hardware streams to exploit tile-level pipelining.
For a given GNN model, the compiler generates the functions in three steps, as illustrated in \Fig{fig:compiling}.

\paragraph{Step 1: Constructing the \ir{} with graph semantics.}
The compiler extracts a generalized GNN computational graph to capture the nature of the given GNN model.
Specifically, it first acquires the raw computational graph from the standard DNN programming frameworks such as TensorFlow and PyTorch, then defuses and replaces the library-customized \ttt{GOP}s into atomic ones (i.e., \bench{Scatter}s, \bench{Gather}s) 
We achieve that by maintaining a list of operations according to the library implementations, for example the \texttt{apply\_edges} / \texttt{update\_all} in DGL and the \ttt{Scatter} in PyG.

Afterward, the compiler splits the computational graph into multiple segments as the graph-native IR to reveal the graph semantics.
Because the tensor types (e.g., tensor for edge or vertex) are changed only by the \bench{GOP}s, the compiler simply splits the model by replacing each \texttt{GOP} with a pair of IR communicational operations \bench{send} and \bench{recv} as annotated by the red solid boxes in \Fig{fig:compiling}.
The output is multiple disconnected GNN model segments, and the compiler then labels the segments with graph semantics based on the communicational operations.
For example, a segment containing \bench{sendOutEdge} will be labeled as \textit{IR.v.x} where the \textit{x} denotes the segment index.
The compiler finally maps the rest computational operations to the corresponding single-item operations (i.e., operations for a single edge or vertex).
It also inserts input and output markers for the whole IR as the entry and exit indicators to finish the IR construction.

\paragraph{Step 2: Optimizing the structure of the \ir{}.}
The proposed graph-native GNN IR not only supports the existing optimizations in deep learning, but also enables the GNN-specific optimizations.
The node and edge in our IR segments represent the operator and feature embedding, which is highly similar to the deep learning computational graph.
So the existing optimizations in deep learning, e.g., subgraph substitution and operator fusion, are fully applicable to our IR.
Besides, the supplementary of the graph semantics enables the space to explore the GNN-specific optimizations, which involves both the computational graph and the input graph. 
We further propose such an optimization based on our IR and we detail it in \Sec{sec:ir-optimization}.

{\renewcommand{\arraystretch}{1.2}
\begin{table*}[t]
\caption{The details of ISA for the \proj{} architecture.}
\label{tbl:isa}
\centering\selectfont
\resizebox{0.9\linewidth}{!}{
\begin{tabular}{ccll}
\toprule
\multicolumn{2}{c}{\textbf{Instruction Type}}    & \multicolumn{1}{c}{\textbf{Examples}}                                                                                                             & \multicolumn{1}{c}{\textbf{Operands}}                                                                      \\ \midrule
\makecell[r]{\multirow{6}{*}{Computational}}     & ELW        & \begin{tabular}[c]{@{}l@{}}arithmetics (\ttt{ADD}, \ttt{SUB}, \ttt{MUL}, \ttt{DIV}), special functions (\ttt{EXP}, \ttt{RELU}),\\ matrix-vector multiplication (\ttt{GEMV})\end{tabular}   & \begin{tabular}[c]{@{}l@{}}data dimensions,  source and destination addresses \\ of embedding memory\end{tabular}     \\ \cmidrule{2-4} 
                                    & GEMM       & \begin{tabular}[c]{@{}l@{}}general matrix multiplication (\ttt{GEMM}), index-guided batched matrix \\ multiplication (\ttt{BMM})   \end{tabular}                     &   \multirow{3}{*}{\begin{tabular}[c]{@{}l@{}} tile id, data dimensions, source and destination \\ addresses of embedding memory\end{tabular} }    \\ \cmidrule{2-3}
                                    & GOP        & \begin{tabular}[c]{@{}l@{}}edge gather (\ttt{GTHR.DST.SUM}, \ttt{GTHR.DST.MAX}), vertex scatter \\ (\ttt{SCTR.OUTE}, \ttt{SCTR.INE})  \end{tabular}                              &              \\ \midrule
\multicolumn{2}{c}{Data-Transfer}                & \begin{tabular}[c]{@{}l@{}}off-chip memory load (\ttt{LD.DST}, \ttt{LD.SRC}, \ttt{LD.EDGE}), store (\ttt{ST.DST})\end{tabular}                                & \begin{tabular}[c]{@{}l@{}}tile id, data dimensions, destination address of \\ embedding memory\end{tabular}                \\ \midrule
\multicolumn{2}{c}{Synchronization}              & \begin{tabular}[c]{@{}l@{}}stream wakeup (\ttt{SIGNAL.E}), fetch new tile / partition ID (\ttt{FCH.TILE}, \\ \ttt{FCH.PTT}), update (\ttt{UPD.PTT}), check (\ttt{CHK.PTT})\end{tabular}                                                & stream ID                                                                                                  \\ \bottomrule
\end{tabular}
}
\end{table*}
}

\paragraph{Step 3: Generating \func{} code.}
Before translating the optimized IR into functions of the instruction sequences, we first adapt the proposed general IR according to the low-level hardware execution model.
In this paper, we target the tiling-based execution model (\Sec{subsec:multistream}), so we need independent and separate \tsfunc{} and \pdfunc{} for the processing of source and destination vertices under a tile and a partition, respectively.
We obtain the functions by further dividing the vertex segments into source and destination parts as illustrated in \Fig{fig:compiling}c.
The compiler first replicates the vertex segments, and then prunes the operations unrelated to the source or destination semantics from each replica.

Given the adapted IR, we generate the instruction \func{}s using the hardware ISA that we describe later.
The source, destination, and edge segments target the \tsfunc{}, \pdfunc{}, and \tefunc{}, respectively.
For each segment, we sort it topologically from the input markers.
The input and output markers are translated into data-transfer instructions; 
the computational and \texttt{send} operations corresponds to the computational instructions;
the \texttt{recv} operations are regarded as barriers for the synchronization instructions to ensure the multi-stream execution semantics.

Since the proposed IR can be multiple disconnected segments as in \Fig{fig:compiling}, the sorting can be deadlocked, which implicates the interaction between the segments.
For the deadlock in a destination segment, the compiler inserts the \ttt{update}, \ttt{signal}, and \ttt{wait} instructions, and resumes the sorting by removing the \texttt{recv} operations in the next source or destination segment; 
for the deadlock at the end of a source segment, the compiler inserts the \ttt{signal} and \ttt{wait} instructions, and resumes the sorting by removing the \texttt{recv}s in an edge segment;
for the deadlock at the end of an edge segment, the compiler checks whether there are still unvisited segments: 
if true, it simply removes the \texttt{recv}s in one unvisited segment and resumes the sorting, or it ends the IR translation to finish the function generation.

\paragraph{Instruction Set Architecture (ISA).}
We propose \proj{} ISA with three types of instructions: computational, data-transfer, and synchronization instructions as shown in \Tbl{tbl:isa}. 
The \textbf{\textit{computational}} instructions are supposed to cover all the operations of \texttt{GEMM}, \texttt{ELW} and \texttt{GOP} appeared in the model computation.
Each of the instructions is coarse-grained and operates on all the edges or vertices in a tile to improve the performance.
The \texttt{ELW} instructions also have matrix and vector versions aiming at different use cases.
The \textbf{\textit{data-transfer}} instructions are designed for loading and storing the embeddings of the edges and vertices.
They are also coarse-grained in the tile level and can be further divided into multiple off-chip memory transactions according to the protocol.
The \textbf{\textit{synchronization}} instructions perform inter-stream synchronization and ensure the correct execution order.

\subsection{IR-Based Compiling Optimization}\label{sec:ir-optimization}

The graph-native IR also provides a wide optimization space for GNN model in addition to generating the low-level code.
Optimizations from both traditional deep learning and graph processing such as graph substitution and operation fusion, forming GNN-specific optimizations, can be automatically applied by compilers through the IR. 
We propose \textit{edge-to-vertex (E2V)} optimization to demonstrate how the IR can be used for GNN model optimization.

The idea of E2V is to move an operation on edge to the vertex if the operation input only involves the source or destination of the edges.
For example, we can move the two matrix-vector multiplications (\texttt{MV}s) to the vertex segment in \Fig{fig:compiling}b as the edge-to-vertex optimization.
Because the nature of one vertex relates to multiple edges, the data scattered to edges will be the same as the source or destination vertex data.
If now an operation applied to the edges uses only the source or destination data, the results would also be the same.
So this operation causes redundant computation.

To detect and eliminate such redundancy, we apply the operation before the data are scattered to the edges.
We first detect the redundancy by traversing and examining the operations in the edge segment from the \ttt{recv}s. 
Operations that use only the source or destination are enqueued.
When an operation does not satisfy the condition, we move the operations in the queue ahead of the corresponding \ttt{send} in the vertex segment.
Finally, we insert extra \ttt{send-recv} pairs for the moved operations whose results are still used in the edge segment, and restart to examine the next \ttt{recv}.
In fact, the E2V optimization can be also applied to original computational graphs, but it may complicate the implementation and incur several unnecessary traversals on the whole graph.

\section{Hardware Architecture}
\label{sec:hardware}

To support the diverse primitive operations and inter-tile pipelining in GNN computation, we design a flexible hardware substrate, as shown in \Fig{fig:overview}.
Generally, we deploy two types of computing units for the primitive operations and launch multiple streams concurrently for the inter-tile pipelining due to the interleaving nature of operations which uses different computation resources in GNN computation.
The streams are created and maintained by a hardware scheduler, which feeds the instructions of the ready stream to the downstream dispatcher.
The dispatcher decodes and issues the incoming instruction to the target components for the execution.
We also design a tile hub and a unified memory for storing the edge list and embeddings of the tiles being processed by the streams.

\subsection{Hardware Component}

In the \proj{} architecture, we deploy a set of computing units and various memory structures.
We later perform a design space exploration (i.e., the number of computing units and memory structure parameters) for justifying our choices.

\paragraph{Computing Unit.} Two types of computing units are deployed for different primitive operations (\Sec{subsec:characterization}): Matrix Unit (MU) and Vector Unit (VU).
The MU is a single systolic array including a weight buffer for the \ttt{GEMM}s. It executes with output stationary dataflow~\cite{Eyeriss} where the input embeddings and weight are feed to the unit at the same time.
The VU is a group of single-instruction-multiple-data (SIMD) cores for the \ttt{ELW}s and \ttt{GOP}s.
The reason for offloading \ttt{GOP}s to the VU instead of another dedicated component is that the atomic operations in \ttt{GOP} are also element-wise with only the operands determined by the tile edge list.
For executing the \ttt{GOP}s, each core is responsible for scattering or gathering one vertex in the tile at a time and fetches the corresponding part of the edge list.
Both the MUs and VUs can be instantiated to have multiple instances to increase parallelism.

\paragraph{Memory.} The on-chip memory consists of two parts: 1) a large unified embedding memory (UEM) for storing the input, intermediate and output embeddings of edges and vertices, and 2) a small tile hub (TH) for the graph tiles containing the edge list and other metadata such as the edge and vertex numbers of the tile.
Owing to the large size of embeddings, we use eDRAM as the UEM. The eDRAM has multiple banks and connects directly with all the computing units to support the multi-streamed parallel execution.
For the TH, we use a small dedicated on-chip SRAM, since the edge lists are accessed more frequently and randomly.
The new data will be loaded to the TH when an eStream finishes tile.
For the interaction between the on-chip and off-chip memory, we design a memory controller that responds to the data-transfer instructions, where the vertex (tile) request is converted to the off-chip memory transactions according to the vertex ID and embedding size (the address and size of the previous tile). 

\subsection{Scheduling}

The \proj{} hardware uses a two-level scheduling, which exploits both the tile-level parallelism (TLP) and the operator-level parallelism (OLP) with streams.
Recall that a stream executes the SDE functions generated from the GNN model on a tile.
We implement it as a group of registers that represent its state.

The first \textit{scheduler} creates and manages the streams for the pipelined execution. 
It adopts a simple first-ready-first-serve policy for scheduling different streams.
When a stream is ready, the scheduler fetches and feeds the current instruction to the input queue of the next level dispatcher, and switches stream state to \textit{issued}.
If the dispatcher queue is full, the schedule will stall.

The second \textit{dispatcher} is responsible for decoding and issuing the incoming instructions from the scheduler.
It also bookkeeps the state of each computing unit.
For a computational instruction, the dispatcher will find a target unit that is ready to issue the instruction. 
If all the target units are busy, the instruction will be added to an instruction queue until the previous instructions and a target unit finish.
We set the queue size to the maximum stream number to avoid congestion.
For a data-transfer and synchronization instruction, the dispatcher will issue the instruction back to the scheduler for state update and to the memory controller for transaction generation, respectively.

\section{Evaluation}\label{sec:evaluation}

In this section, we present the evaluation results for \proj{}. 
We first explain the evaluation methodology and then present the detailed performance results.

\subsection{Methodology}\label{sec:eval:methodology}

\paragraph{Benchmark Datasets and Models.}
The details of the selected datasets can be found in \Tbl{tbl:benchmark:dataset}.
We also select five popular and diverse GNN models for the evaluation.
\begin{itemize}[leftmargin=10pt]
    \vspace*{-0.1cm}
    \item \bench{GCN}~\cite{GCN} is the most famous but simplest GNN model that bridges the gap between spectral- and spatial-based approaches. It consists only of a pair of Scatter-Gather (also known as a \ttt{SpMM}) and a \ttt{GEMM}. 
    \vspace*{-0.2cm}
    \item \bench{GAT}~\cite{GAT} extends the GCN with a multi-head attention mechanism, involving multiple ELWs and GOPs. We only use one head in our evaluation for simplicity.
    \vspace*{-0.2cm}
    \item \bench{SAGE}~\cite{GraphSAGE} generalizes GCN with different aggregator, from which we choose \texttt{maxpool} in our benchmark.
    \vspace*{-0.2cm}
    \item \bench{GGNN}~\cite{GG-NN} employs a gated recurrent unit (GRU). We implement the GRU with separate \ttt{ELW}s and \ttt{GEMM}s on \proj{}, but leverage the GRUcell kernel on GPU.  
    \vspace*{-0.2cm}
    \item \bench{R-GCN}~\cite{RGCN} introduces the GCN to the graph with multiple edge types, which correspond to different weights in the \ttt{GEMM}. We set the type number to 3 and randomly generate the edge type for each benchmark graph.  
    \vspace*{-0.2cm}
\end{itemize}
We run the forward pass of a single layer of each model under a typical 128 for the input and output embedding sizes in all experiments, but it is also straightforward for our accelerator to run different layers and embedding sizes.

{\renewcommand{\arraystretch}{1.2}
\begin{table}[h]
\caption{Details of the graph datasets~\cite{Gunrock} for evaluation.}
\label{tbl:benchmark:dataset}
\centering
\resizebox{\linewidth}{!}{
\begin{tabular}{cccc}\toprule
    \textbf{Dataset} & \textbf{\#Vertex} & \textbf{\#Edge} & \textbf{Type} \\
    \midrule
    ak2010 (AK)           & 45,293     & 108,549    & Redistrict Set            \\
    coAuthorsDBLP (AD)    & 299,068    & 977,676    & Citation Networks         \\  
    hollywood-2009 (HW)   & 1,139,905  & 57,515,616 & Collaboration Networks    \\ 
    cit-Patents (CP)      & 3,774,768  & 16,518,948 & Patent Networks           \\  
    soc-LiveJournal1 (SL) & 4,847,571  & 43,369,619 & Social Networks           \\
    europe-osm (EO)       & 50,912,018 & 54,054,660 & Street Networks        \\ 
    \bottomrule             
\end{tabular}
}
\end{table}
}

\paragraph{System Configuration.}
\Tbl{tbl:configuration} shows the system configuration, where the baseline is DGL~0.5 on one NVIDIA V100 GPU with 32GB memory and two Intel Xeon E5-2630 v4 with 256GB memory, respectively.
For \proj{}, we use a $32\times128$ systolic array as a MU, and eight 32-wide as a VU.
Intuitively, we deploy one dStream, four sStreams and four eStreams on top of one MU and two VUs in \proj{}.
We explore the design space of the streams and computing units later in \Sec{subsec:eval:opt}.

{\renewcommand{\arraystretch}{1.2}
\begin{table}[!htbp]
	\caption{System configurations.}\label{tbl:configuration}
	\centering
	\resizebox{0.48\textwidth}{!}{
		\begin{tabular}{cccc}
			\toprule
			&  \textbf{DGL (CPU)} & \textbf{DGL (GPU)} & \textbf{\PROJ{}}            
			\\ \midrule       
			\begin{tabular}[c]{@{}c@{}}\textbf{Compute Unit}\end{tabular}    
			& \begin{tabular}[c]{@{}c@{}}1.25GHz\\5120 cores\end{tabular}       
			& \begin{tabular}[c]{@{}c@{}}2.20GHz\\20 cores\end{tabular}       
			& \begin{tabular}[c]{@{}c@{}}1GHz~~1$\times$MU (32$\times$128 systolic array), \\ 2$\times$VUs (each with 8$\times$SIMD32 cores) \end{tabular}  
			\\ \midrule  
			\begin{tabular}[c]{@{}c@{}}\textbf{On-Chip Mem.}\end{tabular}    
			& 50MB 
			& 34MB 
			& \begin{tabular}[c]{@{}c@{}}21MB (EM) + 256KB (GU) eDRAM \end{tabular}   \\ \midrule  
			\begin{tabular}[c]{@{}c@{}}\textbf{Off-Chip Mem.}\end{tabular}  
			& \begin{tabular}[c]{@{}c@{}}136GB/s\\DDR4\end{tabular} 
			& \begin{tabular}[c]{@{}c@{}}900GB/s\\HBM-2.0\end{tabular} 
			& \begin{tabular}[c]{@{}c@{}}256GB/s HBM-1.0\end{tabular}  \\ \bottomrule
		\end{tabular}      
	}
\end{table}}

\paragraph{Performance Simulation.}
We develop a C++-based cycle-level architecture simulator to evaluate our system and optimizations.
The simulator also integrates Ramulator~\cite{Ramulator} to simulate the interaction behavior between the off-chip High Bandwidth Memory (HBM) and the on-chip memory.
We validate the performance against HyGCN~\cite{HyGCN} with our best effort, and the functionality of each operation and the tiling-based execution against DGL.

\paragraph{Energy Estimation.}
We estimate the energy consumption of \proj{} with three main components: the multiply-accumulate (MAC) units, the on-chip memory, and off-chip memory. 
For the MAC energy, we obtain the average energy of a single MAC operation by synthesizing a small systolic array with registers under TSMC 16 nm technology and multiplying it with the number of total MAC operations.
We employ Cacti 6.5~\cite{cacti} to measure the dynamic and leakage energy of the two on-chip memory under 32 nm technology and then convert them down to 16 nm~\cite{area_convert}.
The off-chip memory access energy is estimated with 7 pJ/bit~\cite{GraphDynS}.

\paragraph{Area Measurement.}
We mainly focus on the area estimation of on-chip memory and compute units, which is obtained also by the synthesis and Cacti 6.5 as described above.
The schedule and control logic in the \proj{} is based on state machines with simple lookup tables, which is insignificant compared to the compute units and memory.

\begin{figure}[t]
    \centering
    \includegraphics*[width=\linewidth]{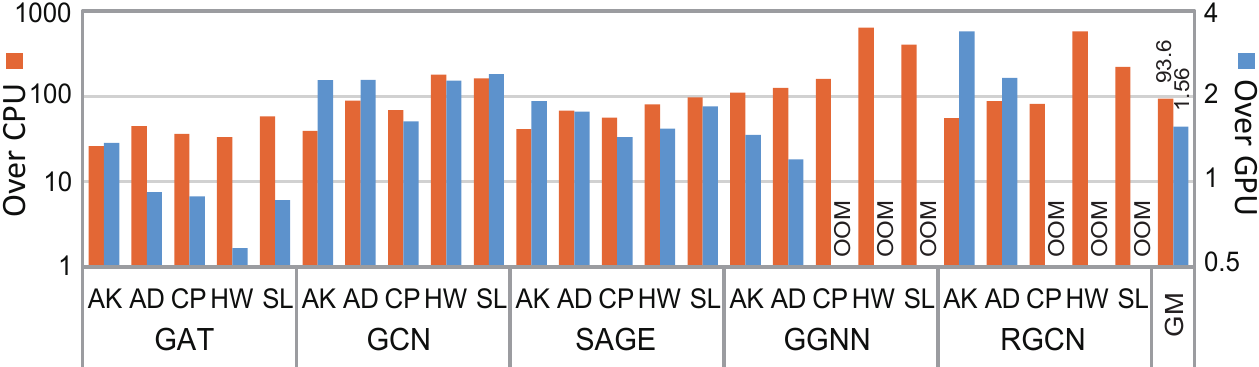}\\
    \vspace{-8pt}
    \caption{\proj{} speedup over the baseline CPU and GPU.}
    \label{fig:eval:performance}
     \vspace{-8pt}
\end{figure}

\begin{figure}[b]
    \centering
    \vspace{-5pt}
    \includegraphics*[width=\linewidth]{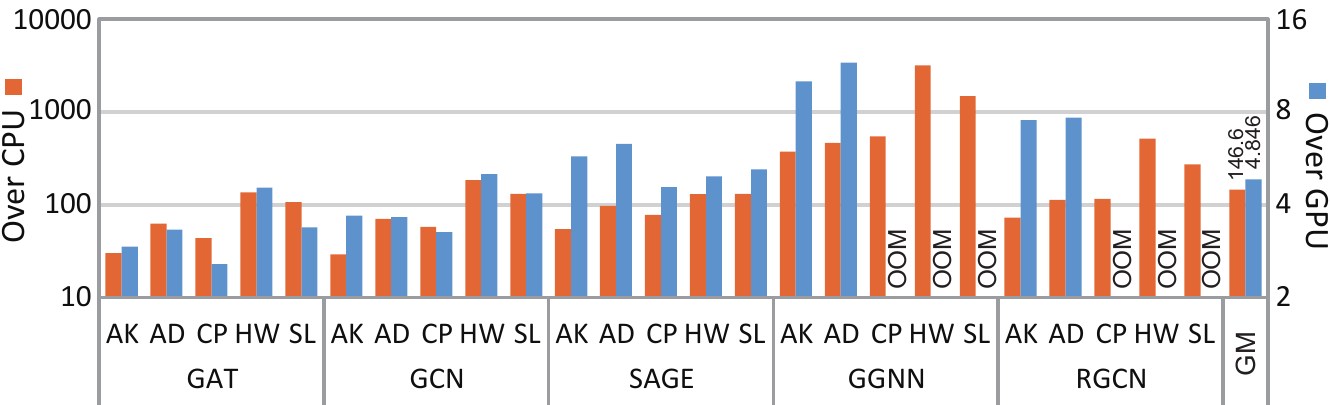}\\
    \caption{\proj{} energy reduction over the baselines.}
    \label{fig:eval:energy}
\end{figure}

\subsection{Detailed Results}\label{subsec:eval:results}

\paragraph{Performance.}
\Fig{fig:eval:performance} shows the performance speedup over the baselines.
We observe that \proj{} outperforms the CPU and GPU, achieve $93.6\times$ and $1.56\times$ speedup on average, respectively.
In the meantime, \proj{} is able to process the large graph datasets and achieves the highest speedup over CPU because of the application of graph tiling.
In contrast, GPU processing the graph as a whole is limited to its off-chip memory size and issues the out-of-memory error. 
We also observe limited speedup and even slowdown for GAT. 
This is because DGL has their special operation support for the softmax attention in the GAT while we implement it with ordinary instructions.
Besides, the dataset HW is much denser than others, so there is less sparsity to exploit and makes the sparse tiling strategy less effective.
Nevertheless, \proj{} still achieves considerable speedups on the other four models.

\paragraph{Energy.}
\Fig{fig:eval:energy} shows the total energy reduction of \proj{} over the baseline CPU and GPU.
We observe that the energy consumptions of the CPU and GPU are $147\times$ and $4.85\times$ that of \proj{} on average, respectively. 
This is not surprising since we leverage multiple dedicated computing units for the different GNN primitives, while the GPU and CPU spend excessive hardware resources to flexibly support various workloads.
Besides, the sparse tiling and reordering also reduce a large number of redundant accesses to both the on-chip and off-chip memory, which takes a large part of the energy consumption.  

\paragraph{Area.}
\Tbl{tbl:eval:area} presents the area breakdown of the \proj{} architecture except for HBM.
The overall area of \proj{} is $53.58~mm^2$, which is down to $6.57\%$ of the baseline GPU die size.
The on-chip memory (including the unified embedding memory and the tile hub) consumes $97.91\%$ area of the \proj{}, and the computing units (including one MU and two VUs) consume about $2.09\%$ area.

{\renewcommand{\arraystretch}{1.2}
\begin{table}[h]
\caption{The area of \proj{} architecture.}
\label{tbl:eval:area}
\centering
\resizebox{\linewidth}{!}{
\begin{tabular}{cccccc}
\toprule

& One MU & One VU & Embedding Mem. & Tile Hub & ~Total~
\\ \midrule

\textbf{Area ($mm^2$)}   & 1.00  & 0.06  & 52.31  & 0.15   & 53.58  
\\ \midrule

\textbf{Percentage} & 1.86\% & 0.12\%  & 97.63\%  & 0.28\%   & 100\%              
\\ \bottomrule

\end{tabular}}
\vspace{-5pt}
\end{table}}

\begin{figure}[t]
    \centering
    \includegraphics*[trim=0.1cm 0.1cm 0.5cm 0.1cm,width=1.0\linewidth]{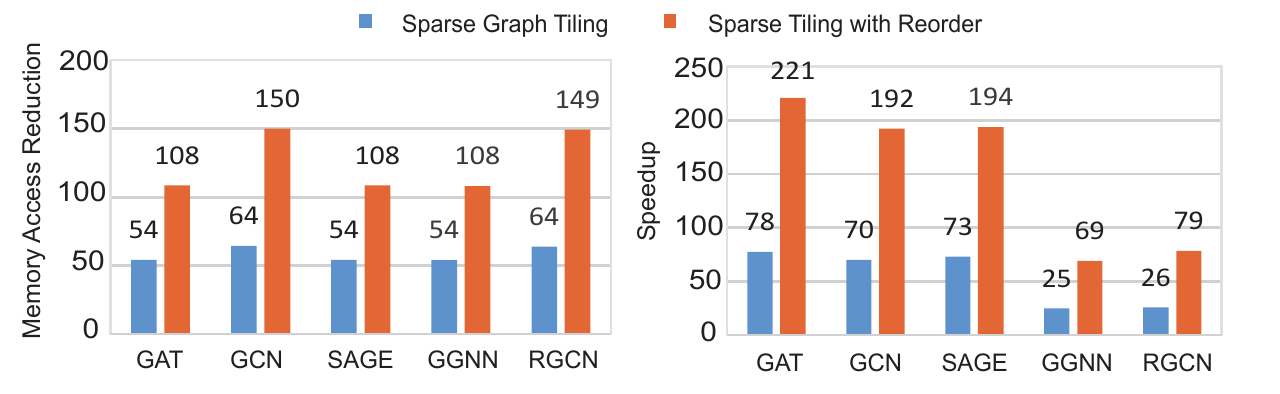}\\
    \caption{The total off-chip memory access reduction (left) and speedup (right) of sparse tiling and reordering over the regular tiling on dataset CP.}
    \label{fig:eval:tiling}
    \vspace{-12pt}
\end{figure}

\subsection{Optimization Effectiveness}\label{subsec:eval:opt}

We analyze the effect of each optimization in \proj{}.

\begin{figure}[b]
    \centering
    \vspace{-8pt}
    \includegraphics*[width=0.99\linewidth]{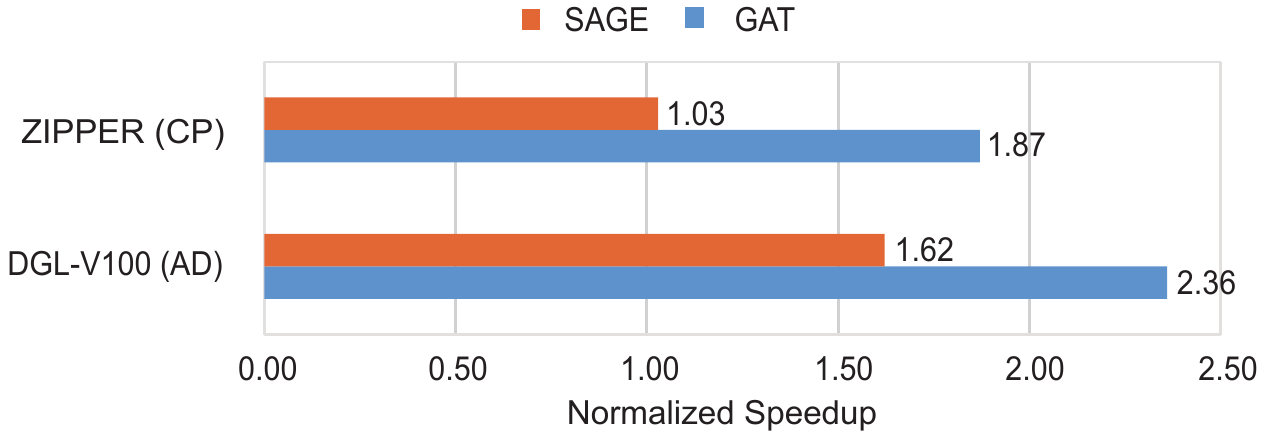}\\
    \vspace{-4pt}
    \caption{Speedup of compiler optimization on dataset CP.}
    \label{fig:eval:cpl}
\end{figure}

\begin{figure*}[t]
    \centering
    \vspace{-4pt}
    \includegraphics*[width=1\linewidth]{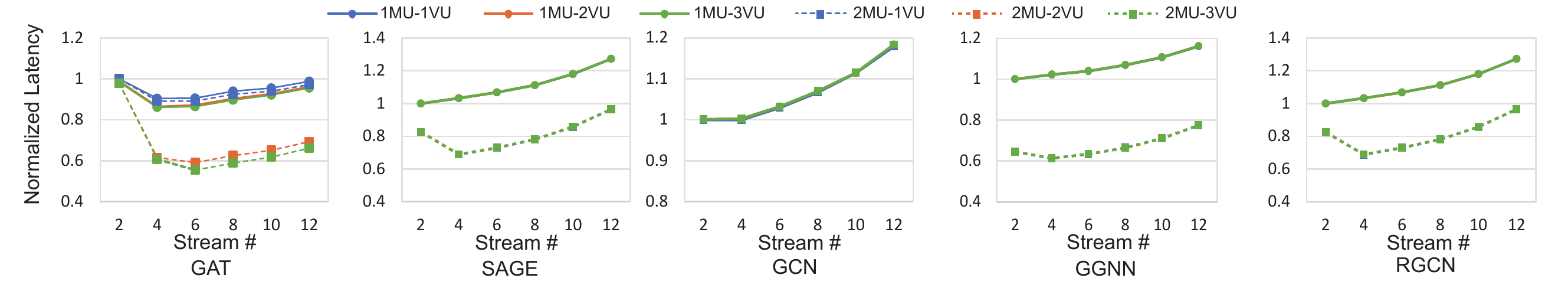}\\
    \vspace{-4pt}
    \caption{\proj{} design space exploration on dataset CP.}
    \label{fig:eval:arch}
    \vspace{-4pt}
\end{figure*}

\paragraph{Sparse Tiling and Reordering.}
We compare the off-chip memory read and execution latency with different tiling strategies: regular tiling, sparse tiling, and sparse tiling with reordering.
\Fig{fig:eval:tiling} shows the results on CP while the results also follow the same trend on other datasets.
We can observe that the two sparse tiling methods provide $58\times$ and $123\times$ of memory access reduction on average.
The lower reduction of \ttt{GAT}, \ttt{SAGE} and \ttt{GGNN} is because they also access the destination vertex embeddings, which cannot be reduced.

Besides, the tiling methods also achieve $48\times$ and $135\times$ speedup on average over the regular tiling.
The reason for the low speedup of \ttt{GGNN} and \ttt{RGCN} is that the two models involve edge-type-guided batch matrix multiplication, which suffers from a longer latency of on-chip memory access and dilutes the benefit of memory access and computation reduction.
But in general, the sparse tiling with reordering can effectively reduce memory access and improve performance.

\paragraph{Compiling Optimization.}
We evaluate the compiling optimization effectiveness by comparing the optimized and the naive implementation of the same GNN models.
\Fig{fig:eval:cpl} shows the speedup results of \texttt{GAT} and \texttt{SAGE}, which we find the opportunities for the compiling optimization.
But libraries have optimized the models manually because of their high popularity. 
So we just implement a naive but straightforward version of the same models using DGL.
\ttt{GAT} and \ttt{SAGE} achieves $1.87\times$ and $1.03\times$ speedup, respectively. 
Our optimization also works on the baseline V100 GPU with DGL and achieves a speedup of $2.36\times$ and $1.62\times$.

\paragraph{Multi-Stream on Hybrid Architecture.}
We evaluate different design choices by changing the numbers of s/eStreams, VU, and MU.
\Fig{fig:eval:arch} shows execution latencies normalized to the result of each model with two s/eStreams, one MU, and two VUs, where we make two observations.
First, there is usually a sweet point of the s/eStream number.
while we increase the s/eStream number, the performance first increases but then decreases with at most $1.72\times$ speedup.
But the sweet points usually vary between models and architectures.
Second, models have different sensitivity to different computational units.
For example, the speedup of \texttt{GAT} changes with both VU and MU, while that of \texttt{SAGE} only changes with MU.
This is because of the different model definition, where the \texttt{SAGE} have much more dense \ttt{GEMM} requiring the MU.

\begin{figure}[t]
    \centering
    \vspace{-4pt}
    \includegraphics*[width=1\linewidth]{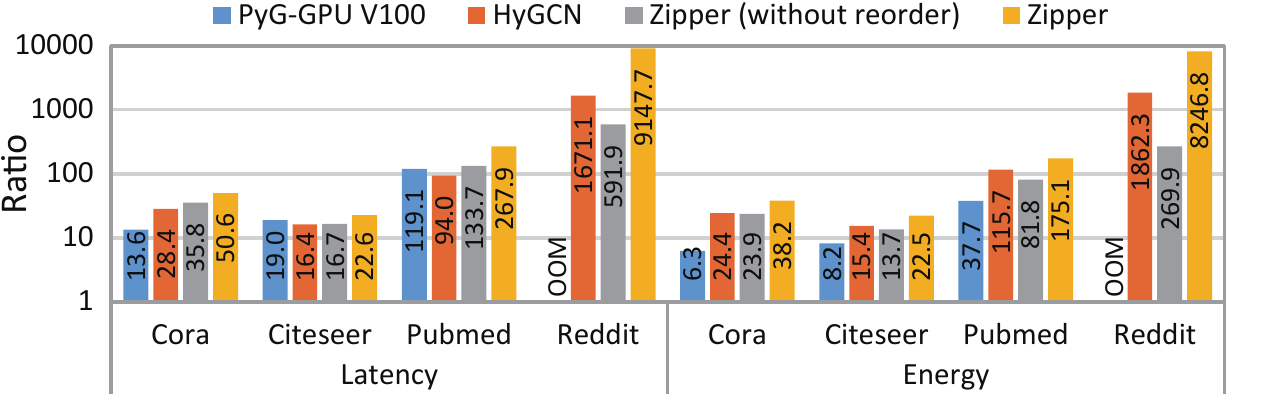}\\
    \vspace{-4pt}
    \caption{Speedup and energy reduction over PyG-CPU.}
    \label{fig:eval:hygcn}
    \vspace{-8pt}
\end{figure}

\subsection{Comparison with HyGCN}

We also compare \proj{} with the current state-of-the-art GCN accelerator HyGCN~\cite{HyGCN}.
To ensure a fair comparison, we use four datasets: Cora, Citeseer, Pubmed and Reddit, and run a full two-layer GCN on \proj{} as described in HyGCN.
We also compare the baseline CPU and GPU with PyG~\cite{PyG} under the same configuration.

\Fig{fig:eval:hygcn} depicts the results.
In the end-to-end comparison, \proj{} outperforms HyGCN in both latency and energy for all the cases. 
We also disable the software reordering to only compare the hardware with HyGCN, and find that \proj{} performs a bit worse than HyGCN but still better than PyG-GPU.
The reason, in addition to the different technology node, is that HyGCN owns a two-stage pipeline specialized for the GCN model, while \proj{} breaks it for better flexibility to support more general GNNs such as GAT and RGCN.

\section{Related Work}\label{sec:related}

The growing scale of graph data and the wide use of GNNs have driven the development of GNN-specific hardware.

\paragraph{GNN Accelerator.}
Most GNN accelerators combine the different components with an efficient but fixed pipeline structure. 
HyGCN~\cite{HyGCN} is the first to design a hybrid architecture with two-stage pipeline for the irregular and regular computation in GNN.
However, it only focuses on the GCN-like models, which are only a small portion of GNNs.
GReTA~\cite{GReTA} and GNNerator~\cite{GNNerator} extend the pipeline with one more component and bidirectional dataflow, respectively.
GraphACT~\cite{GraphACT} and ReGraphX~\cite{ReGraphX} also take CPU and 3D ReRAM techniques into consideration for accelerating the GNN training. 
However, those works are still built or optimized for only a specific class of the real-world GNN models since they do not change the nature of the fixed pipeline structure.

In contract, Auten et al.~\cite{DAC:AutenT020} connect all the components with a crossbar switch in a hardware block to make it possible to support general GNN models from the hardware aspect, which is the closest work to ours.
However, they miss the key information of how a GNN model as well as the input graph is mapped to and executed on the hardware.   
Besides, their distributed block topology potentially limits the DNN component and bandwidth to take their full advantage, and leads to poor performance on large datasets. 
Instead, our architecture supports arbitrary GNN models through the graph-native GNN IR compiler and flexible hardware scheduling.
Meanwhile, our optimizations for the graph tiling and model compiling also improve the overall performance, which is a complete solution for the general GNN acceleration.

\paragraph{SpMM Architectures.}
EnGN~\cite{EnGN} proposes a unified SIMD architecture where they adopt a ring-based dataflow for \ttt{SpMM} to improve hardware utilization. 
AWB-GCN~\cite{AWB-GCN} proposes three auto-tuning techniques based on a special \ttt{SpMM} unit to address the issue of workload imbalance in GNNs.
In a word, these works focus on the irregularity issues in \ttt{Gather} operation, which are inherited from traditional graph processing. 
They are orthogonal to our work since we focus more on the inter-tile pipelining, and are applicable to our Vector Unit for executing the \ttt{GOP}s in the GNN models.

\section{Conclusion}\label{sec:conclusion}

In this work, we propose \proj{}, a general and scalable GNN acceleration system that implements the inter-tile pipelining to exploit the tile- and operation-level parallelism.
We first characterize the GNN computation and identify two main problems: excessive memory footprint and GNN primitive operation interleaving.
These two problems motivate the idea of the inter-tile pipelining.
We first leverage the graph tiling to address the problem of the excessive memory footprint and then pipeline the resulted tiles through a tiling-based multi-stream execution model.
However, because of the graph semantic unawareness of the high-level GNN programming model, which is inherited from the traditional DNN frameworks, it is non-trivial to do the GNN model conversion from the programming model into our execution model.
So we further propose the GNN Intermediate Representation (IR) and the associated compiler to recover the graph semantics and generate the low-level programs automatically.
Finally, we provide architectural support for the tiling-based multi-stream execution, which leads to the \proj{} architecture. 
We also propose two optimizations based on graph tiling and compiling to improve the  
The proposed system achieves $93.6\times$ speedup with $147\times$ energy reduction and $1.56\times$ speedup with $4.85\times$ energy reduction over CPU and GPU solution on average.

\newpage
\bibliographystyle{IEEEtranS}
\bibliography{refs}

\end{document}